\documentclass[10pt,twocolumn,oneside,a4paper,final]{IEEEtran}
\IEEEoverridecommandlockouts
\def\BibTeX{{\rm B\kern-.05em{\sc i\kern-.025em b}\kern-.08em
    T\kern-.1667em\lower.7ex\hbox{E}\kern-.125emX}}
    
\usepackage{amsmath,amssymb,amsfonts,graphicx,mathtools,nccmath,amsthm,float}
\usepackage{multicol,tikz,cite,array,booktabs,url}
\usepackage{cite}
\usepackage{textcomp}
\usepackage{xcolor}
\usepackage{algorithm}
\usepackage{algpseudocode,algorithmicx}
\usepackage{acronym}
\usepackage{xfrac}

\algnewcommand{\LineComment}[1]{\State \(\#\) #1}
\algnewcommand\algorithmicinput{\textbf{Set}}
\algnewcommand\Set{\item[\algorithmicinput]}
\algnewcommand\algorithmicinitial{\textbf{Initialize}}
\algnewcommand\Initialize{\item[\algorithmicinitial]}

\let\oldReturn\Return
\renewcommand{\Return}{\State\oldReturn}

\graphicspath{{figures/}}
\DeclareGraphicsExtensions{.pdf}

\def\endthebibliography{%
  \def\@noitemerr{\@latex@warning{Empty `thebibliography' environment}}%
  \endlist
}

\newcommand{\myVec}[1]{{\boldsymbol{#1}}}
\newcommand{\myMat}[1]{{\boldsymbol{#1}}}
\newcommand{\mySet}[1]{\mathcal{#1}}

\newcommand{\Nusers}{K}
\newcommand{\Nantennas}{N_{\rm R}}
 
\newcommand{\Ncoeffs}{N}

\newcommand{\norm}[1]{\left\lVert#1\right\rVert}

\acrodef{adc}[ADC]{analog-to-digital convertor}
\acrodef{cs}[CS]{compressed sensing}
\acrodef{dtft}[DTFT]{discrete-time Fourier transform}
\acrodef{dnn}[DNN]{deep neural network} 
\acrodef{csi}[CSI]{Channel State Information}
\acrodef{map}[MAP]{maximum a-posteriori probability}
\acrodef{snr}[SNR]{signal-to-noise ratio}
\acrodef{bs}[BS]{Base Station} 
\acrodef{em}[EM]{electromagnetic} 
\acrodef{iot}[IOT]{Interent of Things}
\acrodef{mimo}[MIMO]{Multiple-Input Multiple-Output}
\acrodef{mse}[MSE]{mean-squared error}
\acrodef{pdf}[PDF]{probability density function}
\acrodef{rv}[RV]{random variable}
\acrodef{ml}[ML]{machine learning}
\acrodef{fec}[FEC]{forward error correction}
\acrodef{rs}[RS]{Reed-Solomon}
\acrodef{lti}[LTI]{linear time-invariant}
\acrodef{wss}[WSS]{wide-sense stationary}
\acrodef{psd}[PSD]{Power Spectral Density}
\acrodef{ser}[SER]{symbol error rate} 
\acrodef{ber}[BER]{bit error rate} 
\acrodef{sgd}[SGD]{stochastic gradient descent} 
\acrodef{isi}[ISI]{intersymbol interference}  
\acrodef{awgn}[AWGN]{additive white Gaussian noise} 
\acrodef{ut}[UT]{user terminal} 
\acrodef{mmw}[mmWave]{millimeter wave}
\acrodef{noma}[NOMA]{non-orthognal multiple access}
\acrodef{mac}[MAC]{mulitple access channel}
\acrodef{fl}[FL]{Federated learning}
\acrodef{ris}[RIS]{Reconfigurable Intelligent Surface} 
\acrodef{ofdm}[OFDM]{Orthogonal Frequency Division Multiplexing}

\newtheorem{Lem}{Lemma}

\DeclareMathOperator{\trace}{Tr}

\title{Wideband Multi-User MIMO Communications with Frequency Selective RISs: Element Response Modeling and Sum-Rate Maximization}
\author{\IEEEauthorblockN{Konstantinos D. Katsanos$^1$, Nir Shlezinger$^2$, Mohammadreza F. Imani$^3$, and  George C. Alexandropoulos$^1$}\\
\IEEEauthorblockA{$^1$Department of Informatics and Telecommunications,
National and Kapodistrian University of Athens, Greece\\
$^2$School of Electrical and Computer Engineering, Ben-Gurion University of the Negev, Beer-Sheva, Israel\\
$^3$School of Electrical, Computer, and Energy Engineering, Arizona State University, USA\\
emails: \{kkatsan, alexandg\}@di.uoa.gr, nirshl@bgu.ac.il, mohammadreza.imani@asu.edu}
\thanks{This work has been supported by the EU H2020 RISE-6G project under grant number 101017011.}
\vspace{-0.9cm}
}

\begin{document}
%
\maketitle
\thispagestyle{empty}

\begin{abstract}
Reconfigurable Intelligent Surfaces (RISs) are an emerging technology for future wireless communication systems, enabling improved coverage in an energy efficient manner. RISs are usually metasurfaces, constituting of two-dimensional arrangements of metamaterial elements, whose individual response is commonly modeled in the literature as an adjustable phase shifter. However, this model holds only for narrowband communications, and when wideband transmissions are utilized, one has to account for the frequency selectivity of metamaterials, whose response usually follows a Lorentzian-like profile. In this paper, we consider the uplink of a wideband RIS-empowered multi-user Multiple-Input Multiple-Output (MIMO) wireless system with Orthogonal Frequency Division Multiplexing (OFDM) signaling, while accounting for the frequency selectivity of RISs. In particular, we focus on designing the controllable parameters dictating the Lorentzian response of each RIS metamaterial element, in order to maximize the achievable sum rate. We devise a scheme combining block coordinate descent with penalty dual decomposition to tackle the resulting challenging optimization framework. Our simulation results reveal the achievable rates one can achieve using realistically frequency selective RISs in wideband settings, and quantify the performance loss that occurs when using state-of-the-art methods which assume that the RIS elements behave as frequency-flat phase shifters.\vspace{-0.1cm}
\end{abstract}
\begin{IEEEkeywords}
Reconfigurable intelligent surface, frequency selectivity, multi-user MIMO, rate optimization, wideband systems.\vspace{-0.6cm}
\end{IEEEkeywords}

\section{Introduction}
\label{sec:intro}
\vspace{-0.1cm}
Future generations of wireless communications are subject to a multitude of diverse performance requirements. These include ultra-high data rate, energy efficiency, wide network coverage and connectivity, as well as ultra-low reliability and end-to-end latency \cite{Samsung}. 
One of the main challenges in meeting these demands stems from the difficulty to guarantee coverage when serving multiple users in harsh non-line-of-sight environments, as commonly encountered in various urban setups \cite{chowdhury20206g}. A key emerging technology that is envisioned to enable energy-efficient improved coverage in future wireless communications is the \ac{ris} \cite{huang2019reconfigurable,di2019smart}. Such metasurfaces are capable of realizing reconfigurable reflection patterns, allowing the dynamic control of the propagation of information-bearing electromagnetic waves \cite{dai2020reconfigurable}, thus paving the way to the recent vision of smartly programmable and sustainable wireless environments \cite{rise6g}. 
	
The ability of \acp{ris} to impact radio wave propagation in a power-efficient and low-cost manner gave rise to growing interests in \ac{ris}-empowered communication systems \cite{huang2019reconfigurable,di2019smart}. 
Most communication-oriented studies for such systems assume a simplistic model for the responses of the RIS constituting elements, which are commonly modeled as controllable phase shifters that do not vary with frequency \cite{Zhang2020_OFDM_MIMO, An_2021a, Zhang_2021_wideband}. In practice, however, metamaterial-based \acp{ris} can only realize such reflection responses in relatively narrow bands centralized at a given resonant frequency \cite{DSmith-2017PRA}, and their reflection patterns depends on the incident wave's angle \cite{chen2020angle}. Consequently, when using wideband transmissions, one must account for the frequency selective response of the metamaterial. Very recently, the authors in \cite{Li2021} considered a frequency-dependent model for the amplitude and phase shift variations of the RIS elements, and studied the sum-rate maximization problem with \ac{ofdm}. However, a metamaterial's frequency response usually takes a Lorentzian form \cite{Mancera2017}. While the Lorentzian spectral behavior of metasurfaces in wideband communications was recently considered for their application as active massive \ac{mimo} antennas \cite{shlezinger2020dynamic, wang2020dynamic,Shlezinger2019_DMA}, it has not been studied to date in wireless communication systems with passive metamaterial-based RISs.

Motivated by the above, in this paper, we capitalize on the inherent frequency selectivity of metasurfaces and exploit it for optimizing wideband multi-user MIMO communications with passive RISs. We first introduce a physics-compliant model for the Lorentzian frequency response of the RIS metamaterial elements, which captures the parameters one can externally control to modify the surface's reflection profile. Then, we present an optimization framework for setting the RIS controllable parameters in order to maximize the achievable average sum rate in the uplink of wideband \ac{ris}-empowered multi-user \ac{mimo} systems with \ac{ofdm} signaling. Our algorithm tackles the challenging coupling of the parameters in the Lorentzian response via combining Block Coordinate Descent (BCD) with the Penalty Dual Decomposition (PDD) method \cite{Shi2020_PDD} in an alternating fashion. Our numerical results demonstrate that the proposed design allows to achieve notably improved achievable sum-rate performance compared to using conventional \ac{ris} configuration approaches, which assume elements acting as frequency-flat phase shifters. 
	
Throughout the paper, boldface lower-case and boldface upper-case letters represent vectors and matrices, respectively, while $\mathbf{I}_n$ is the $n\times n$ ($n\geq2$) identity matrix and notation $\mathbf{X} \succeq \mathbf{0}$ indicates that $\mathbf{X}$ is positive semi-definite.
The transpose, Hermitian transpose, trace, determinant, and the real part of a complex quantity are written as $(\cdot)^T$,  $(\cdot)^H$, ${\rm {Tr}}\left(\cdot\right)$, $|\cdot|$, and $\Re\{ \cdot \}$, respectively, while $\mathbb{C}$ is the set of complex numbers, $\jmath\triangleq\sqrt{-1}$ is the imaginary unit, $[\mathbf{X}]_{i,j}$ denotes $\mathbf{X}$'s $(i,j)$-th element, $\operatorname{vec}(\mathbf{X})$ vectorizes $\mathbf{X}$, and 
$\operatorname{vec}_{\rm d}(\mathbf{X})$ denotes the vector obtained by the diagonal elements of the square $\mathbf{X}$.
	

\vspace{-0.3cm}
\section{RIS Response and System Modeling}
\label{sec:Sys_Model}\vspace{-0.1cm}
\begin{figure}
	\centering
		\includegraphics[scale=0.60]{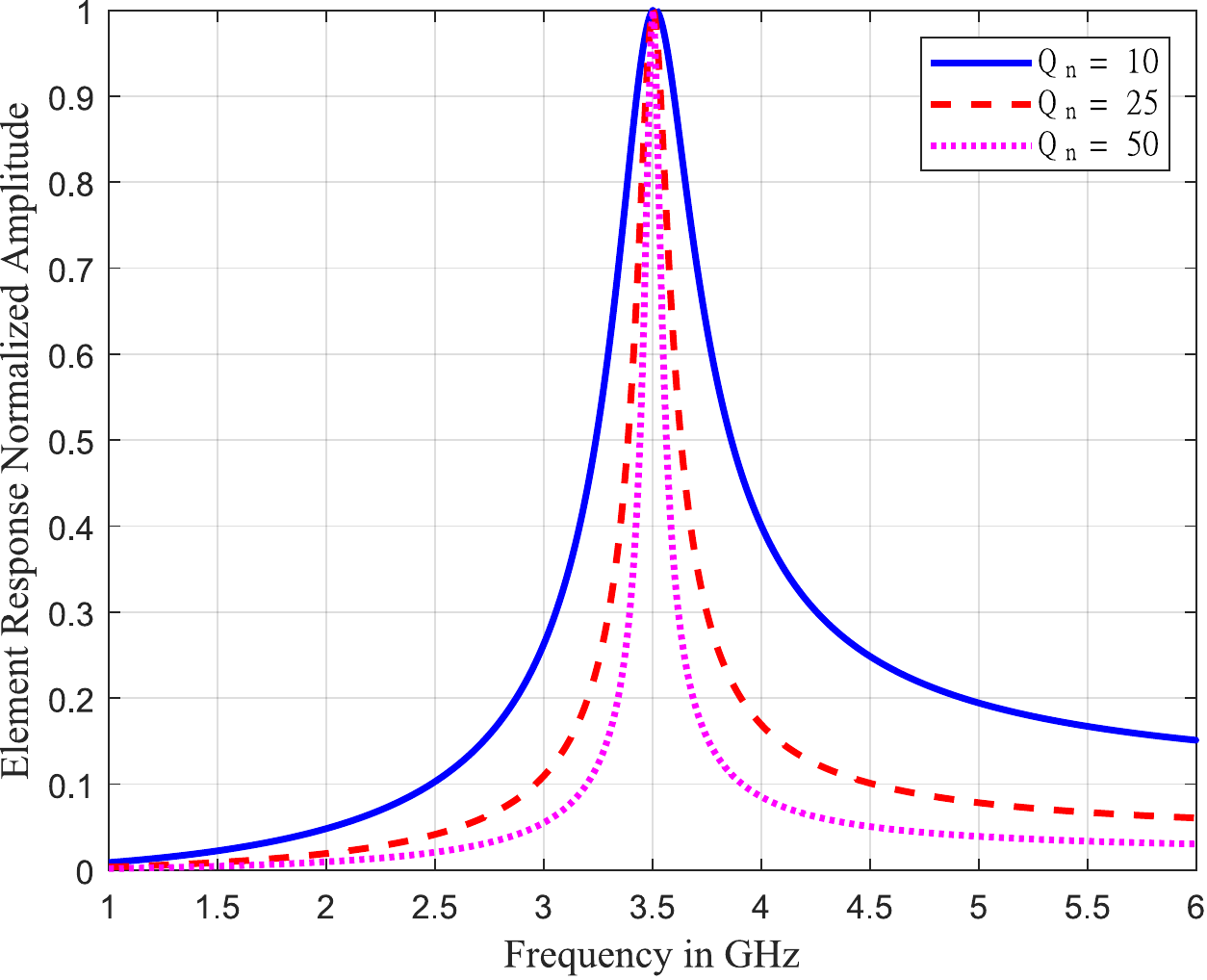}
	\caption{The frequency response of the proposed RIS element model as a function of the frequency in GHz for various values of the quality factor $Q_n$.}\vspace{-0.5cm}
\label{fig:RISDiagram1}
\end{figure}
\subsection{Metamaterial's Frequency Response Model}\label{subsec:RIS_model}\vspace{-0.1cm}
Passive RISs are often realized using metasurfaces \cite{dai2020reconfigurable,rise6g}. Their constituting metamaterials, with externally controllable parameters, enable diverse reflection patterns for various operation objectives \cite{alexandg_2021}.
To model feasible reflection profiles with metamaterial-based RISs (referred to, henceforth, as RISs), we consider a surface with $N$ elements, and let $s_n[i]$ and $r_n[i]$ with $n \in \{1,2,\ldots,N\} \triangleq \mySet{N}$ be the impinging signal at the $n$-th RIS element and its reflected version, respectively, at the discrete time instance $i$. The response of each $n$-th element is typically modeled as a tunable frequency-flat phase shifter, i.e., $r_n[i]\triangleq \phi_n[i]s_n[i]$ with $\phi_n[i]\triangleq e^{\jmath\psi_n[i]}$, where $\psi_n[i] \in [0,2\pi)$ can be configured individually. However, this extensively used model, serves as a narrowband approximation for the RIS element response; in practice, each metamaterial element has been shown to behave as a resonant circuit \cite{DSmith-2017PRA}.

The resonant elements (described as inductor-capacitor resonators in circuit theory) of a metasurface exhibit a frequency dispersive property. 
By assuming electrically small metamaterial elements to implement the RIS, we can model each element as a polarizable dipole, whose frequency response takes the following Lorentzian form:
\begin{equation}
	\label{eqn:Lorentz1}
	\phi_n(\omega) =  \frac{F_n \omega^2}{\omega_n^2 -\omega^2+\jmath\kappa_n \omega},
\end{equation}
where $F_n$, $\omega_n$, and $\kappa_n$ are the element-dependent oscillator strength, angular resonance frequency, and damping factor, respectively, which can be externally controllable. The quality factor $Q_n\triangleq\frac{\omega_n}{\kappa_n}$ determines the bandwidth that an RIS element can influence. Smaller $Q_n$ values result in flatter frequency responses (as commonly considered in the RIS literature), while large $Q_n$'s indicate narrower frequency profiles; this behavior is illustrated in Fig.~\ref{fig:RISDiagram1}. By defining the vectors $\mathbf{r}[i] \triangleq \big[r_1[i]\,r_2[i]\,\cdots\, r_{\Ncoeffs}[i]\big]^T$ and $\mathbf{s}[i] \triangleq \big[s_1[i]\,s_2[i]\,\cdots\,s_{\Ncoeffs}[i]\big]^T$, as well as the diagonal matrix $\myMat{\Phi}[i] \triangleq {\rm diag}\big\{\phi_1[i]\,\phi_2[i]\,\cdots\,\phi_{\Ncoeffs}[i]\big\}$, we can write the output signals at the $\Ncoeffs$ \ac{ris} metamaterial elements as follows:
\begin{equation}\label{eqn:Reflected_Signal}
\mathbf{r}[i] = \myMat{\Phi}[i]\star \mathbf{s}[i],
\end{equation}
where the operand $\star$ stands for the multivariate convolution.

\vspace{-0.4cm}
\subsection{System and Channel Models}\label{subsec:Channel_model}
\begin{figure}
	\centering
		\includegraphics[scale=0.85]{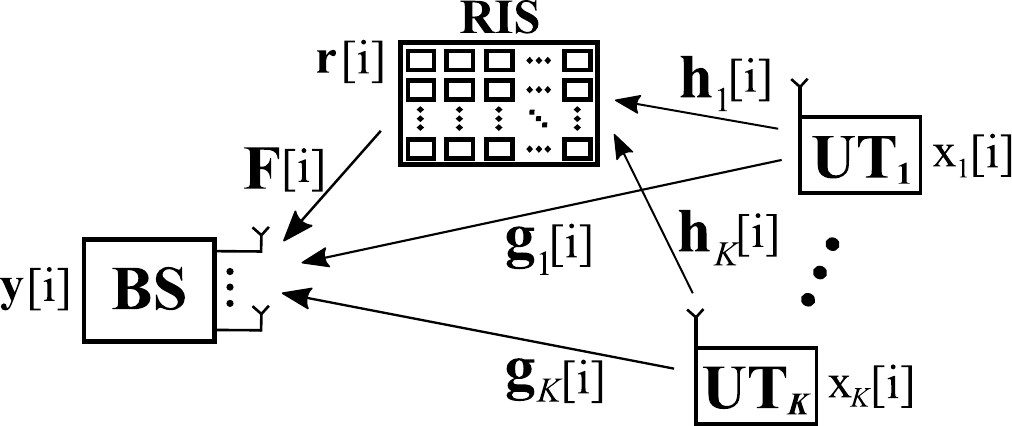}
	\caption{The wideband RIS-empowered MIMO OFDM system with $K$ UTs.}\vspace{-0.5cm}
\label{fig:System_Model}
\end{figure}

\vspace{-0.1cm}
We consider an RIS-empowered multi-user MIMO OFDM system operating in the uplink direction, as illustrated in Fig.~\ref{fig:System_Model}. A \ac{bs} with $\Nantennas$ antennas serves $\Nusers$ single-antenna User Terminals (UTs) with the help of the passive RIS of the previous section. Let $x_k[i]$ be the transmitted signal from the $k$-th UT, with $k=1,2,\ldots,K$, at the time instance $i$. 
We assume frequency selective wireless channels with finite memory of $Q$ time instances. In particular, let $\mathbf{h}_k[i]\in\mathbb{C}^{\Ncoeffs}$ be the multivariate  impulse response modeling the channel between the RIS and the $k$-th UT, while $\mathbf{F}[i]\in \mathbb{C}^{\Nantennas\times \Ncoeffs}$ and $\mathbf{G}[i]\triangleq\big[\mathbf{g}_1[i]\,\mathbf{g}_2[i]\,\cdots\,\mathbf{g}_K[i]\big]\in\mathbb{C}^{\Nantennas\times K}$ (with $\mathbf{g}_k[i]\in \mathbb{C}^{\Nantennas}$ referring to the $k$-th UT) denote the multivariate impulse responses of the RIS-BS channel and the direct channels between the BS and UTs, respectively. The $\Nantennas$-element discrete-time received signal vector at the BS can be expressed as \cite{heathMimo2018}
	 \begin{align} 
	 \mathbf{y}[i] &= \sum_{\tau=0}^{Q-1}\mathbf{F}[\tau] \mathbf{r}[i-\tau] + \sum_{\tau=0}^{Q-1}\mathbf{G}[\tau] \mathbf{x}[i-\tau] + \mathbf{w}[i]\notag \\& 
	 =\mathbf{F}[i] \star \mathbf{r}[i] + \mathbf{G}[i] \star \mathbf{x}[i] + \mathbf{w}[i], 
	 \label{eqn:ChannelIO1}
\end{align}
\!\!\!where $\mathbf{x}[i] \triangleq \big[x_1[i]\,x_2[i]\cdots\, x_{\Nusers}[i]\big]^T$ and $\mathbf{w}[i]$ denotes the zero-mean additive  Gaussian noise vector with covariance matrix $\sigma^2\mathbf{I}_{N_{\rm R}}$. The signal impinging at the RIS is given by 
\begin{equation}
	\label{eqn:Conv2}
	\mathbf{s}[i] = 
	\sum_{k=1}^K \sum_{\tau=0}^{Q-1}\mathbf{h}_k[\tau]x_k[i-\tau],
\end{equation}
yielding, by using \eqref{eqn:Reflected_Signal}, the following reflected signal: 
\begin{align}\label{eqn:RIS1}
	\mathbf{r}[i] = \sum_{k=1}^{\Nusers}\myMat{\Phi}[i]\star \mathbf{h}_k[i] \star x_k[i].
\end{align}
By defining the $\Ncoeffs \times \Nusers$ matrix sequence $\mathbf{J}[i]$ such that $\mathbf{J}[i] \triangleq \big[\mathbf{h}_1[i]\, \mathbf{h}_2[i]\, \cdots\, \mathbf{h}_\Nusers[i] \big]$,  \eqref{eqn:RIS1} can be re-written as $\mathbf{r}[i] = \myMat{\Phi}[i]\star\mathbf{J}[i] \star \mathbf{x}[i]$. Consequently, the channel input-output relationship \eqref{eqn:ChannelIO1} for a fixed \ac{ris} configuration  $\myMat{\Phi}[i]$ is given by
\begin{equation}
    \label{eqn:RIS2}
    	\mathbf{y}[i] \triangleq \left(\mathbf{F}[i] \star \myMat{\Phi}[i]\star\mathbf{J}[i] + \mathbf{G}[i]\right) \star \mathbf{x}[i] + \mathbf{w}[i].
\end{equation}

\vspace{-0.4cm}
\section{Sum-Rate Maximization} \label{sec:RIS_optim}\vspace{-0.1cm}
\subsection{Problem Formulation} \label{subsec:Prob_Form}\vspace{-0.1cm}
We consider the problem of tuning the frequency-selective RIS to optimize the achievable sum-rate performance. To formulate this communication objective, we assume that the UTs utilize wideband modulations, and use $\myMat{C}_{\mathbf{x}}(\omega)$ to denote the \ac{psd} of $\mathbf{x}[i]$. We also assume that the \ac{bs} has full \ac{csi}, i.e., it knows the channel input-output relationship \eqref{eqn:RIS2} and the \ac{psd} of the transmitted signals, which uses to configure the RIS via its controller \cite{rise6g}. Following these assumptions, we can express the achievable sum rate for the considered system as \cite{shlezinger2020dynamic}: 
\begin{equation}
    \tilde{\mathcal{R}} \triangleq \dfrac{1}{2\pi} \int_{0}^{2\pi} \log \left\lvert \mathbf{I}_{N_R} + \sigma^{-2} \mathbf{D}(\omega) \mathbf{D}(\omega)^H  \right\rvert d\omega. \label{eqn:Rate_integral}
\end{equation}
where $\mathbf{D}(\omega) \triangleq \big(\tilde{\mathbf{F}}(\omega) + \mathbf{G}(\omega)\big)\myMat{C}_{\mathbf{x}}^{1/2}(\omega)$ with  $\tilde{\mathbf{F}}(\omega)$ and $\mathbf{G}(\omega)$ being the multivariate discrete-time Fourier transform of $\mathbf{F}[i] \star \mathbf{\Phi}[i]\star\mathbf{J}[i]$ and $\mathbf{G}[i]$, respectively.
The maximization of the achievable sum rate with respect to $\myMat{C}_{\mathbf{x}}(\omega)$ is the sum-capacity, 
which can be approached using multi-carrier modulations, i.e., \ac{ofdm}. By dividing the spectrum into $B$ finite frequency bins, the sum-rate performance in \eqref{eqn:Rate_integral} can be approximated as follows: 
\begin{equation} \label{eqn:RateExpression_2}
\mathcal{R} \triangleq \dfrac{1}{B} \sum_{b=1}^B \log \left\lvert \mathbf{I}_{N_{\rm R}} + \sigma^{-2} \mathbf{D}(\omega_b) \mathbf{D}(\omega_b)^H \right\rvert.
\end{equation}

The latter expression depends on the \ac{ris} configuration $\{\mathbf{\Phi}(\omega_b)\}_{b = 1}^B$, which is encapsulated in $\{\mathbf{D}(\omega_b)\}_{b = 1}^B$. 
We next write the sum rate in \eqref{eqn:RateExpression_2} as $\mathcal{R}(\tilde{\myVec{\Phi}})$ with $\tilde{\myVec{\Phi}} \triangleq [\boldsymbol{\varphi}_1\,\boldsymbol{\varphi}_2\,\cdots\,\boldsymbol{\varphi}_B] \in \mathbb{C}^{N\times B}$, where $\boldsymbol{\varphi}_b\triangleq[\phi_1(\omega_b)\,\phi_2(\omega_b)\,\cdots\,\phi_N(\omega_b)]^T$ is obtained from \eqref{eqn:Lorentz1} for a given $\omega_b$. As discussed in Subsection~\ref{subsec:RIS_model}, the response of the \ac{ris} metamaterial elements is dictated by the parameters $\{F_n, \omega_n, \kappa_n\}_{n=1}^N$. Consequently, the configuration of the \ac{ris} that maximizes the achievable sum-rate per $\omega_b$ is the solution of the following optimization problem:
\begin{align*}
\begin{split}
    \mathcal{OP}: \,\, \max_{\{F_n, \omega_n, \kappa_n\}_{n=1}^N} \quad& \mathcal{R}(\tilde{\myVec{\Phi}}) \\
    \text{s.t.} \quad \quad& [\tilde{\myVec{\Phi}}]_{n,b} = \dfrac{F_n \omega_b^2}{\omega_n^2 - \omega_b^2 + \jmath\kappa_n \omega_b},\\
    \quad \quad& \lvert [\tilde{\myVec{\Phi}}]_{n,b} \rvert \leq 1, \forall n \in \mathcal{N}, \forall b \in \mathcal{B},
\end{split}
\end{align*}
where $\mathcal{B}\triangleq \{1,2,\ldots B\}$. In the sequel, we present an algorithm for designing the parameters of the considered frequency selective \ac{ris}.

\vspace{-0.4cm}
\subsection{Proposed RIS Design Solution} \label{subsec:RIS_Optimization}\vspace{-0.1cm}
To tackle the non-linear objective function and the non-convex constraints in $\mathcal{OP}$, we first transform the former into a more tractable form, based on the following Lemma \cite{Shi2011_WMMSE}:
\begin{Lem} \label{Lemma_WMMSE}
Suppose that $\mathbf{M} \in \mathbb{C}^{n \times n}$ with $\mathbf{M}\succeq\mathbf{0}$ is defined~as:
\begin{equation} \label{eqn:MSE_matrix}
\mathbf{M} = \left(\mathbf{U}^H \mathbf{B}\mathbf{C} - \mathbf{I}_n\right) \left(\mathbf{U}^H \mathbf{B}\mathbf{C} - \mathbf{I}_n\right)^H + \mathbf{U}^H\mathbf{R}\mathbf{U},
\end{equation}
for some $\mathbf{U}, \mathbf{B} \in \mathbb{C}^{m \times n}$, $\mathbf{C} \in \mathbb{C}^{n \times n}$, and $\mathbf{R} \in \mathbb{C}^{m \times m}$ with $\mathbf{R}\succ\mathbf{0}$. Let also the scalar function $f(\mathbf{S},\mathbf{U})\triangleq \log\lvert\mathbf{S}\rvert - \trace(\mathbf{S}\mathbf{M}) + \trace(\mathbf{I}_n)$ with $\mathbf{S} \in \mathbb{C}^{n \times n}$. It then holds that:
\begin{equation} \label{eqn:lemma-WMMSE}
 \log \left\lvert \mathbf{I}_n + (\mathbf{B} \mathbf{C})^H \mathbf{R}^{-1} \mathbf{B} \mathbf{C} \right\rvert = \max_{\mathbf{U}, \mathbf{S} \succ \mathbf{0}} f(\mathbf{S},\mathbf{U}),
\end{equation}
where  \eqref{eqn:lemma-WMMSE} is maximized by setting  $\mathbf{S}_{\rm opt} = \mathbf{M}^{-1}$.
\end{Lem}

By introducing the auxiliary variables\footnote{From now on, subscript $b$ denotes dependence on the frequency bin $\omega_b$.} $\mathbf{S}_b \in \mathbb{C}^{K \times K}$ and $\mathbf{U}_b \in \mathbb{C}^{N_{\rm R} \times K }$ $\forall b \in \mathcal{B}$, and defining the mean squared error matrix $\mathbf{M}_b \triangleq \left( \mathbf{U}_b^H \mathbf{D}_b - \mathbf{I}_K \right) \left( \mathbf{U}_b^H \mathbf{D}_b - \mathbf{I}_K \right)^H + \sigma^2 \mathbf{U}_b^H \mathbf{U}_b$, the sum rate in \eqref{eqn:RateExpression_2} can be equivalently expressed as follows:
\begin{equation}
\label{eqn:WMMSE_Rate_Transf}
	\mathcal{R} = \max_{\{\mathbf{U}_b, \mathbf{S}_b \succ \mathbf{0}\}, \tilde{\boldsymbol{\phi}}} \quad \sum_{b=1}^B \left( \log \lvert \mathbf{S}_b \rvert - \trace(\mathbf{S}_b \mathbf{M}_b) + K \right),
\end{equation}
where Lemma \ref{Lemma_WMMSE} is applied for each summand in \eqref{eqn:RateExpression_2} and the factor $B$ is omitted without loss of optimality. Even with the transformed objective function \eqref{eqn:WMMSE_Rate_Transf}, $\mathcal{OP}$ is still non-convex due to its coupled variables. However, it can be shown to be convex with respect to each block of variables, namely $\{\mathbf{U}_b, \mathbf{S}_b \}$, for a fixed $\tilde{\boldsymbol{\phi}}\triangleq{\rm vec}(\tilde{\myVec{\Phi}})$. Therefore, to optimize the auxiliary matrix variables, we adopt a BCD approach, as presented in the sequel.

\subsubsection{Optimizing w.r.t. $\{\mathbf{U}_b\}$}
By fixing $\mathbf{S}_b$ and $\tilde{\boldsymbol{\phi}}$, applying some algebraic manipulations, and omitting constant terms, we arrive at the following expression for the optimum $\mathbf{U}_b$ in \eqref{eqn:WMMSE_Rate_Transf}:
\begin{align}
  \mathbf{U}_b^{\rm opt} =& \mathop{\arg\max}\limits_{\mathbf{U}_b} \,\, -\trace(\mathbf{S}_b \mathbf{U}_b^H \mathbf{D}_b \mathbf{D}_b^H \mathbf{U}_b) + \trace(\mathbf{S}_b \mathbf{U}_b^H \mathbf{D}_b) \notag \\ 
 	& \qquad +\trace(\mathbf{S}_b \mathbf{D}_b^H \mathbf{U}_b) - \trace(\mathbf{S}_b) - \sigma^2 \trace({\mathbf{S}_b \mathbf{U}_b^H \mathbf{U}_b}) \notag \\
 	\stackrel{(a)}{=}& \,\, (\sigma^2 \mathbf{I}_{N_R} + \mathbf{D}_b \mathbf{D}_b^H)^{-1} \mathbf{D}_b,  \label{eqn:U_i_opt}
\end{align}
where $(a)$ is obtained by comparing the derivative to zero. 

\subsubsection{Optimizing w.r.t. $\{\mathbf{S}_b\}$}
According to Lemma \ref{Lemma_WMMSE}, $\mathbf{S}_b = \mathbf{M}_b^{-1}$, so it suffices to substitute $\mathbf{U}_b^{\rm opt}$ in the expression of $\mathbf{M}_b$ and take the inverse. After some straightforward manipulations and by invoking the matrix inversion lemma, it follows~that:
\begin{equation} \label{eqn:S_i_opt}
 \mathbf{S}_b^{\rm opt} = \mathbf{I}_K + \sigma^{-2} \mathbf{D}_b^H \mathbf{D}_b.
\end{equation}

\subsubsection{Optimizing w.r.t. $\tilde{\{\boldsymbol{\phi}\}}$}
After some algebraic manipulations on the equivalent rate expression in \eqref{eqn:WMMSE_Rate_Transf}, it can be shown that each of its $B$ terms, as a function of $\boldsymbol{\Phi}_b$, is equal to:
\begin{equation} \label{eqn:rate_reformulation}
\begin{aligned}
	\mathcal{R}_b = &-\trace(\boldsymbol{\Phi}_b^H \mathbf{R}_{1,b} \boldsymbol{\Phi}_b \mathbf{R}_{2,b}) -\trace(\boldsymbol{\Phi}_b (\mathbf{R}_{3,b} - \mathbf{R}_{4,b})) \\ 
	&-\trace(\boldsymbol{\Phi}_b^H (\mathbf{R}_{3,b} - \mathbf{R}_{4,b} )^H) + c_b,
\end{aligned}
\end{equation}
where $\mathbf{R}_{1,b} \triangleq \mathbf{F}_b^H \mathbf{U}_b \mathbf{S}_b \mathbf{U}_b^H \mathbf{F}_b$, $\mathbf{R}_{2,b} \triangleq \mathbf{J}_b \mathbf{J}_b^H$, $\mathbf{R}_{3,b} \triangleq \mathbf{J}_b \mathbf{G}_b^H \mathbf{U}_b \mathbf{S}_b \mathbf{U}_b^H \mathbf{F}_b$, $\mathbf{R}_{4,b} \triangleq \mathbf{J}_b \mathbf{S}_b \mathbf{U}_b^H \mathbf{F}_b$, 
and the constant terms $c_b \triangleq -\trace(\mathbf{S}_b \mathbf{U}_b^H \mathbf{G}_b \mathbf{G}_b^H \mathbf{U}_b) + \trace(\mathbf{S}_b \mathbf{U}_b^H \mathbf{G}_b) + \trace(\mathbf{S}_b \mathbf{G}_b^H \mathbf{U}_b) - \trace(\mathbf{S}_b) - \sigma^2 \trace(\mathbf{S}_b \mathbf{U}_b^H \mathbf{U}_b)$.
Then, by defining $\mathbf{A}_b \triangleq \mathbf{R}_{1,b} \odot \mathbf{R}_{2,b}^T$, $\mathbf{b}_b \triangleq \operatorname{vec}_{\rm d}(\mathbf{R}_{3,b} - \mathbf{R}_{4,b})$, and utilizing the identities \cite[Th. 1.11]{Zhang_2017}, the sum-rate optimization problem in the Lorentzian parameters can be re-written as follows:
\begin{align*}
\begin{split}
    \mathcal{OP}': \,\, \min_{\{F_n, \omega_n, \kappa_n\}_{n=1}^N} \quad& \tilde{\boldsymbol{\phi}}^H \tilde{\mathbf{A}} \tilde{\boldsymbol{\phi}} + 2\Re \left\{ \tilde{\boldsymbol{\phi}}^H \tilde{\mathbf{b}}^* \right\} - \tilde{c} \\
    \text{s.t.} \quad \quad& [\tilde{\myVec{\Phi}}]_{n,b} = \dfrac{F_n \omega_b^2}{\omega_n^2 - \omega_b^2 + \jmath\kappa_n \omega_b},\\
    \quad \quad& \lvert [\tilde{\myVec{\Phi}}]_{n,b} \rvert \leq 1, \forall n \in \mathcal{N}, \forall b \in \mathcal{B},
\end{split}
\end{align*}
where $\tilde{\mathbf{A}}\triangleq \operatorname{blkdiag}\{ \mathbf{A}_b \}_{b = 1}^B$, $\tilde{\mathbf{b}} \triangleq [\mathbf{b}_1^T,\mathbf{b}_2^T,\dots,\mathbf{b}_B^T]^T$, and $\tilde{c} \triangleq \sum_{b=1}^B c_b$. 

The reformulated optimization problem $\mathcal{OP}'$ is simpler than $\mathcal{OP}$, because its objective function is a convex quadratic function with respect to $\tilde{\boldsymbol{\phi}}$. The convexity is justified by observing that $\tilde{\mathbf{A}} \succeq \mathbf{0}$, since it is the Hadamard product of positive semi-definite matrices \cite[Th. 1.9]{Zhang_2017}. On the other hand, the objective needs to be optimized with respect to the Lorentzian parameters, which dictate $\tilde{\boldsymbol{\phi}}$ in a non-linear manner, based on the Lorentzian form in \eqref{eqn:Lorentz1}. Thus, to solve $\mathcal{OP}'$, we adopt the PDD method \cite{Shi2020_PDD}, which is an Augmented Lagrangian (AL) method. This method is suitable for $\mathcal{OP}'$ because its equality and inequality constraints can be separated into two sub-problems, which can be solved iteratively until convergence. Capitalizing on the PDD method, with penalty parameter $\rho$, the $\mathcal{OP}'$'s AL problem is expressed~as
\begin{align*}
\begin{split}
\mathcal{OP}'_{\rm AL}: \,\, \min_{\tilde{\boldsymbol{\phi}},\{F_n, \omega_n, \kappa_n\}_{n=1}^N} \quad&
  g \triangleq \tilde{\boldsymbol{\phi}}^H \tilde{\mathbf{A}} \tilde{\boldsymbol{\phi}} + 2\Re \left\{ \tilde{\boldsymbol{\phi}}^H \tilde{\mathbf{b}}^* \right\}  - \tilde{c}\\
 & \!\!\!\!+\!\dfrac{1}{2\rho} \norm{\tilde{\boldsymbol{\phi}} \!-\! \tilde{\mathbf{d}}\left(\{F_n, \omega_n, \kappa_n\}\right) \!+\! \rho \boldsymbol{\lambda} }^2  \\
\text{s.t.} \quad \quad &\lvert [\tilde{\myVec{\Phi}}]_{n,b} \rvert \leq 1, \forall n \in \mathcal{N}, \forall b \in \mathcal{B},
\end{split}
\end{align*}
where $\tilde{\mathbf{d}} \in \mathbb{C}^{NB}$ is the vector corresponding to the equality constraint of $\mathcal{OP}'$ and $\boldsymbol{\lambda}$ denotes the dual variable vector associated with the equality constraint $\tilde{\boldsymbol{\phi}} = \tilde{\mathbf{d}}$. Then, $\mathcal{OP}'_{\rm AL}$ can be solved based on a two-layer iteration, with the inner layer alternatively updating $\tilde{\boldsymbol{\phi}}$ and the desired Lorentzian parameters, while the outer layer updates $\rho$ and $\boldsymbol{\lambda}$.

\begin{algorithm}[!t]
\begin{algorithmic}[1]
\caption{Proposed PDD-based Method Solving $\mathcal{OP}'_{\rm AL}$}
\label{alg:OP_AL_PDD_Algorithm}
\State \textbf{Input:} $p_{\rm out}$, $\epsilon_{\rm in} > 0$, $\epsilon_{\rm out} > 0$, $\tilde{\mathbf{A}}$, $\tilde{\mathbf{b}}$, as well as feasible $\tilde{\boldsymbol{\phi}}^{(0)}$, $F_n^{(0)}$, $\omega_n^{(0)}$, $\kappa_n^{(0)}$, $\boldsymbol{\lambda}^{(0)}$, $\rho$, and $\mu$.
\For{$ p_{\rm out} = 1,2,\dots$}
	\State Set the inner iteration number $p_{\rm in}=0$.
    \For{$ p_{\rm in} = 1,2,\dots$}
        \State Compute $\tilde{\boldsymbol{\phi}}^{(p_{\rm in})}$ solving $\mathcal{OP}'_{\rm AL, \tilde{\boldsymbol{\phi}}}$ via PG,
        
        \hspace{0.35cm} according to \eqref{eqn:PG_step_1} and \eqref{eqn:PG_step_2}.
        \State Compute $\left\{\hat{F}_n^{(p_{\rm in})}, \hat{\omega}_n^{(p_{\rm in})}, \hat{\kappa}_n^{(p_{\rm in})}\right\}_{n=1}^N$ solving
        
        \hspace{0.35cm} \eqref{eqn:lorentzian_update_ls} via Levenberg-Marquardt.
        \If $\left\lvert \sfrac{\left(g\left(\tilde{\boldsymbol{\phi}}^{(p_{\rm in})}\right) - g\left(\tilde{\boldsymbol{\phi}}^{(p_{\rm in} - 1)} \right) \right)}{g\left(\tilde{\boldsymbol{\phi}}^{(p_{\rm in})}\right)} \right\rvert \leq \epsilon_{\rm in}$, 
        \State\textbf{break};		
		\EndIf
	\EndFor
	\State Set $\tilde{\boldsymbol{\phi}}^{(p_{\rm out})} = \tilde{\boldsymbol{\phi}}^{(p_{\rm in})}$.
	\State Set $\tilde{\mathbf{d}}^{(p_{\rm out})}$ using \eqref{eqn:Lorentz1} with $\big\{\hat{F}_n^{(p_{\rm in})}, \hat{\omega}_n^{(p_{\rm in})}, \hat{\kappa}_n^{(p_{\rm in})}\big\}$.
	\State Set $\boldsymbol{\lambda}^{(p_{\rm out})} = \boldsymbol{\lambda}^{(p_{\rm out}-1)} + \rho^{-1}\left(\tilde{\boldsymbol{\phi}}^{(p_{\rm out})} - \tilde{\mathbf{d}}^{(p_{\rm out})}\right)$.
	\State Compute the decreased penalty parameter  $\rho \leftarrow \mu \rho$.
	\If $\norm{\tilde{\boldsymbol{\phi}}^{(p_{\rm out})} - \tilde{\mathbf{d}}^{(p_{\rm out})}}_{\infty} \leq \epsilon_{\rm out}$,
	\State\textbf{break};	
	\EndIf
\EndFor
\State \textbf{Output:} $F_n^{(p_{\rm out})}$, $\omega_n^{(p_{\rm out})}$, and $\kappa_n^{(p_{\rm out})}$ $\forall$$n\in\mathcal{N}$.
\end{algorithmic}
\end{algorithm}

In the inner layer, whose first task is to optimize $\mathcal{OP}'_{\rm AL}$ with respect to $\tilde{\boldsymbol{\phi}}$, without restricting its elements to exhibit the Lorentzian response, the resulting problem is expressed as:
\begin{align*}
\begin{split}
\mathcal{OP}'_{\rm AL, \tilde{\boldsymbol{\phi}}}: \,\, \min_{\tilde{\boldsymbol{\phi}}} \quad g \quad \text{s.t.} \quad \lvert [\tilde{\myVec{\Phi}}]_{n,b} \rvert \leq 1, \forall n \in \mathcal{N}, \forall b \in \mathcal{B},
\end{split}
\end{align*}
which is convex, since the objective is the sum of a convex quadratic plus a norm function under convex constraints. To solve it, we propose a Projected Gradient (PG) approach. In particular, at each $\ell$-th iteration step with $\ell = 0,1,\dots$, we consider the following steps:
\begin{align}
    \mathbf{x}^{(\ell)} &= \operatorname{Proj}_{\mathcal{F}} \left( \tilde{\boldsymbol{\phi}}^{(\ell)} - \beta\, \nabla_{\tilde{\boldsymbol{\phi}}} g\left(\tilde{\boldsymbol{\phi}}^{(\ell)}\right) \right), \label{eqn:PG_step_1}  \\
    \tilde{\boldsymbol{\phi}}^{(\ell+1)} &= \tilde{\boldsymbol{\phi}}^{(\ell)} + \alpha^{(\ell)}\left( \mathbf{x}^{(\ell)} - \tilde{\boldsymbol{\phi}}^{(\ell)} \right) \label{eqn:PG_step_2},
\end{align}
where $\beta > 0$, $\alpha^{(\ell)} \in (0,1]$ is a step size chosen according to the Armijo rule along the feasible direction \cite{bertsekas1999nonlinear}, and $\operatorname{Proj}_{\mathcal{F}}$ denotes the element-wise projection operator onto the feasible set $\mathcal{F}$ of $\mathcal{OP}'_{\rm AL, \tilde{\boldsymbol{\phi}}}$, according to which, if the argument's modulus is greater than one, it is normalized. 

\begin{algorithm}[!t]
\begin{algorithmic}[1]
\caption{Proposed BCD Method Solving $\mathcal{OP}$}
\label{alg:OP_overall_Algorithm}
\State \textbf{Input:} $m\!=\!0$, $\epsilon\!>\!0$, $M_{\max}$, and feasible $\{F_n^{(0)},\omega_n^{(0)},\kappa_n^{(0)}\}$. 
\For{$ m = 1,2,\dots,M_{\max}$}	
	\For{$b = 1,2,\dots,B$}
		\State Compute $\mathbf{D}_b^{(m)} = \mathbf{G}_b + \mathbf{F}_b \boldsymbol{\Phi}_b^{(m)} \mathbf{J}_b$.
		\State Compute $\mathbf{U}_b^{(m)}$ $\forall b \in \mathcal{B}$ according to \eqref{eqn:U_i_opt}.
		\State Compute $\mathbf{S}_b^{(m)}$ $\forall b \in \mathcal{B}$ according to \eqref{eqn:S_i_opt}.
		\State Compute $\mathbf{A}_b^{(m)} = \mathbf{R}_{1,b}^{(m)} \odot (\mathbf{R}_{2,b}^T)^{(m)}$, 
	
		\hspace{0.35cm}$\mathbf{b}_b^{(m)} \triangleq \operatorname{vec}_{\rm d}(\mathbf{R}_{3,b}^{(m)} - \mathbf{R}_{4,b}^{(m)})$, and $c_b$.
	\EndFor
	\State Compute 
	$\{F_n^{(m)}, \omega_n^{(m)},\kappa_n^{(m)}\}$ using Algorithm \ref{alg:OP_AL_PDD_Algorithm}.
	\If $\left\lvert\left(\mathcal{R}^{(m)} - \mathcal{R}^{(m-1)}\right)/\mathcal{R}^{(m)}\right\rvert \leq \epsilon$ 
		\State\textbf{break};
	\EndIf
\EndFor
\State \textbf{Output:} $F_n^{(m)}$, $\omega_n^{(m)}$, and $\kappa_n^{(m)}$ $\forall n \in \mathcal{N}$.
\end{algorithmic}
\end{algorithm}

Then, in the second step of the inner layer, we optimize $\mathcal{OP}'_{\rm AL}$ with respect to $\{F_n, \omega_n, \kappa_n\}_{n=1}^N$. This results in the following simplified sub-problem, after eliminating the irrelevant terms and constants and letting $\mathbf{f} \triangleq \tilde{\boldsymbol{\phi}} + \rho \boldsymbol{\lambda}$:
\begin{equation} \label{eqn:lorentzian_update_ls}
\{\hat{F}_n, \hat{\omega}_n, \hat{\kappa}_n\}_{n=1}^N=
\mathop{\arg\min}\limits_{\{F_n, \omega_n, \kappa_n\}_{n=1}^N} \norm{\mathbf{f} - \tilde{\mathbf{d}}(\{F_n, \omega_n, \kappa_n\})}^2,
\end{equation}
where the elements of $\tilde{\mathbf{d}}$ take the Lorentzian form \eqref{eqn:Lorentz1}. This reduced problem belongs to the family of non-linear least-squares problems, especially when dealing with the angular frequency $\omega_n$ and the damping factor $\kappa_n$. To solve this problem, we employ the Levenberg-Marquardt algorithm. 

For the outer layer of the proposed PDD-based algorithm, the dual variable $\boldsymbol{\lambda}$ is updated by $\boldsymbol{\lambda} = \boldsymbol{\lambda} + \rho^{-1}(\tilde{\boldsymbol{\phi}} - \tilde{\mathbf{d}})$.
Concerning the penalty parameter $\rho$, it is updated by multiplying it with a constant scaling factor $\mu < 1$, i.e., $\rho \leftarrow \mu \rho$, which is used to force the equality constraint to be approached during the subsequent iterations. All above steps solving $\mathcal{OP}'_{\rm AL}$ via PDD are summarized in Algorithm~\ref{alg:OP_AL_PDD_Algorithm}. Finally, Algorithm~\ref{alg:OP_overall_Algorithm} includes all steps for solving $\mathcal{OP}$'s sum-rate maximization.

The computational complexity of the proposed algorithms is analyzed based on their algorithmic steps, as follows. In Algorithm~\ref{alg:OP_AL_PDD_Algorithm}, the worst case complexity results from the PG method and the Levenberg-Marquardt algorithm, i.e., Steps 5 and 6, respectively. In particular, for the PG method, the required computational cost is $\mathcal{O}(I_{\rm PG}(BN)^{1.5})$ due to the gradient computation of function $g$ \cite{Shewchuk_1994}, with $I_{\rm PG}$ denoting the iterations needed for convergence. In addition, the Levenberg-Marquardt algorithm requires $\mathcal{O}((3N)^3)$ computations \cite{nocedal1999numerical}, which is justified by taking into account that for each RIS element three parameters are optimized. Then, the total complexity of Algorithm~\ref{alg:OP_AL_PDD_Algorithm} is $C_{\rm PDD} = \mathcal{O}(I_{\rm out} I_{\rm in} (I_{\rm PG}(BN)^{1.5} + (3N)^3))$, where $I_{\rm out}$ and $I_{\rm in}$ express the numbers of outer and inner PDD iterations, respectively. Similarly, the complexity of the overall Algorithm~\ref{alg:OP_overall_Algorithm} is given by $C_{\mathcal{OP}} = \mathcal{O}(B M_{\max} (N_{\rm R}^3+K^3 + C_{\rm PDD}) )$, due to the matrix inversions in Steps 5 and 6.

\vspace{-0.6cm}
\subsection{Discussion}\vspace{-0.1cm}
The proposed Algorithm~\ref{alg:OP_overall_Algorithm} enables to optimize metamaterial-based \acp{ris} in a manner that is aware of their frequency selectivity. This affects the RIS phase profiles when applied to wideband signals. Our numerical evaluations, reported in the following Section~\ref{sec:Num_Eval}, demonstrate that our approach yields higher rates compared to utilizing conventional \ac{ris} configuration methods, designed assuming that the RIS elements behave as controllable phase shifters in wideband communications, where the frequency selectivity of the \ac{ris} is not negligible. 

Our design is particularly tailored to control the parameters of each metamaterial element that dictates its frequency response. In practice, one is often interested in restricting the search space of these parameters. For instance, the damping factor is related to the quality factor of the element \cite{DSmith-2017PRA}, which is typically challenging to tune to levels beyond some threshold.  An additional parameter which is known to affect the reflection pattern of \acp{ris} is the incident angle \cite{chen2020angle}, which can be incorporated into the channel model. Nonetheless, we leave the adaptation of our design algorithms to account for the effect of different incident angles and constrained parameter spaces for future work. Furthermore, while our approach is suitable to tackle $\mathcal{OP}$, there is no guarantee that it yields the optimal setting due to the non-convex nature of the problem. In such cases, one can consider data-driven optimization based on the proposed algorithm, via, e.g., the learn-to-optimize framework \cite{zappone2019wireless} or by unfolding the iterative optimization into a neural network \cite{shlezinger2020model}. Moreover, Algorithm~\ref{alg:OP_overall_Algorithm} requires full \ac{csi}, which is often challenging to acquire in \ac{ris}-empowered communications \cite{alexandropoulos2021hybrid}, motivating its possible combination with learning-based \ac{csi}-agnostic joint \ac{ris}-receiver settings, as proposed in \cite{Wang2021jointly}. We leave these extensions to future research. 

\vspace{-0.3cm}
\section{Numerical Results}\vspace{-0.1cm}
\label{sec:Num_Eval}

\begin{figure}[!t]
	\centering
		\includegraphics[scale=0.55]{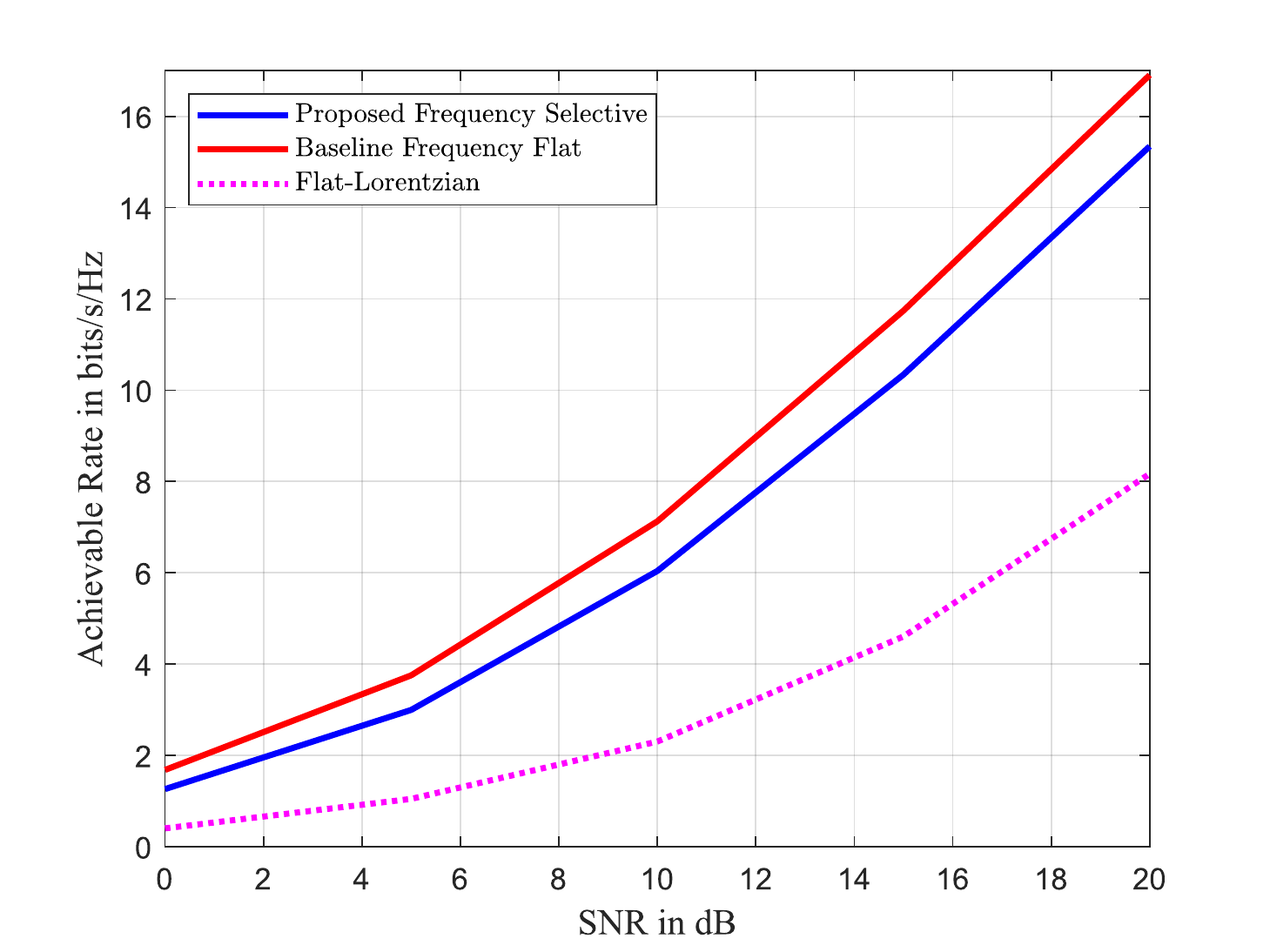}
	\caption{Achievable sum-rate in bps/Hz as a function of the \ac{snr} in dB for the proposed design of a $20$-element frequency-selective RIS for the wideband uplink communication with $4$ UTs, $\Nantennas = 8$, $B = 16$, and $Q = 4$.}\vspace{-0.5cm}
\label{fig:Rate_vs_SNR}
\end{figure}
\subsection{Experimental Setup}\vspace{-0.1cm}
In this section, we investigate the performance of our proposed algorithm in configuring for frequency-selective \acp{ris} in order to maximize the achievable sum-rate performance. We consider Rayleigh fading channels for $\{\mathbf{F}[i]\}_{i=0}^{Q-1}$, $\{\mathbf{J}[i]\}_{i=0}^{Q-1}$, and $\{\mathbf{G}[i]\}_{i=0}^{Q-1}$, with $Q$ being the number of delayed taps in the time-domain impulse response for each link. Each channel is assumed to consist of independent random entries with zero mean and unit variance, multiplied by distance dependent pathloss. The pathloss values are set with the exponent $2.2$ for the \ac{ris}-involved channels (i.e., $\mathbf{F}$ and $\mathbf{J}$), and with $3$ for the direct channel $\mathbf{G}$. In our simulations, the \ac{bs} is located in the origin of the $xy$ plane, whereas the UTs lie on a circle area of radius $3m$ and center at the position $(15m,17.5m)$, with fixed positions that were randomly generated. The \ac{ris} is placed at the position with coordinates $(15m,12.5m)$. In addition, we considered $\Nusers = 4$ users, $\Nantennas = 8$ \ac{bs} antennas, $B = 16$ frequency bins, and $Q = 4$ taps. The Lorentzian parameters were constrained as: $F_n \in (0,1]$, $\omega_n > 0 $ and $\lvert \kappa_n \rvert \leq 100$, while $\lvert \omega \rvert \leq \pi$. The achievable sum-rate performance results were averaged over $100$ independent Monte Carlo realizations.

The convergence thresholds for the proposed algorithms were set as $\epsilon_{\rm in} = 10^{-4}$, $\epsilon_{\rm out} = 10^{-5}$, and $\epsilon = 10^{-3}$, while $\mu$ was selected in the interval $(0.7,0.99)$. For comparison purposes, we have implemented the frequency-flat \ac{ris} phase-shifting algorithm proposed in \cite[Sec. IV]{Zhang2020_OFDM_MIMO}. We evaluated the resulting \ac{ris} configuration benchmark twice: once when the resulting configuration tunes the \ac{ris} reflection pattern at the central frequency (for both magnitude and phase), while at the remaining frequencies the \ac{ris} takes the wideband Lorentzian form. This benchmark, which represents the case where one tunes the \ac{ris} for wideband signals using conventional settings designed for frequency-flat \acp{ris}, is termed as "Flat-Lorentzian." In addition, we also computed as "Baseline Frequency Flat" the rate achieved when the \ac{ris} is indeed frequency flat, representing the rate one expects to achieve using narrowband design approaches. Note that those rates are not actually achievable due to the frequency selectivity under wideband transmissions. 

\begin{figure}[!t]
	\centering
		\includegraphics[scale=0.55]{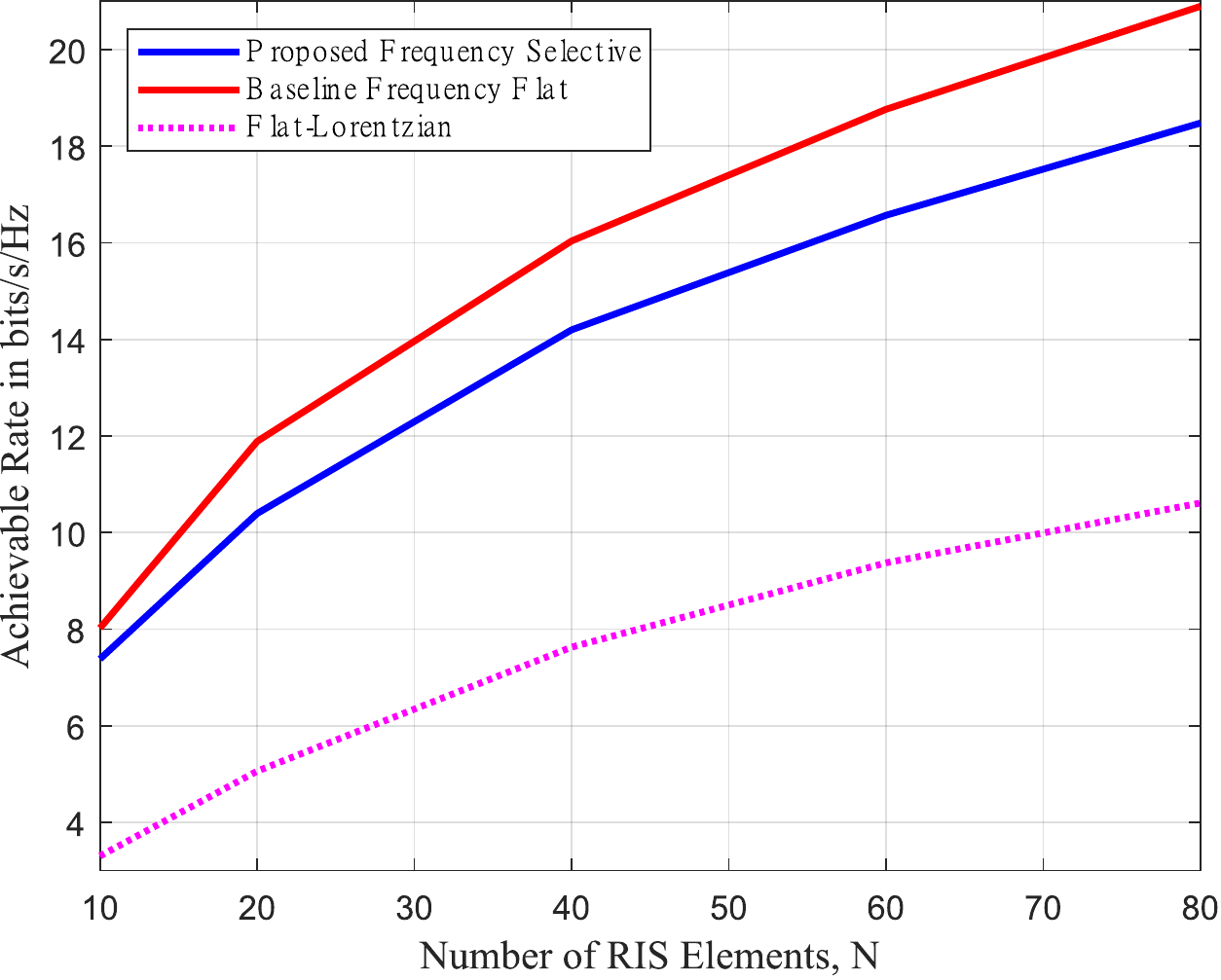}
	\caption{Achievable sum rates in bps/Hz versus the number $N$ of the \ac{ris} metamaterial elements with the proposed design for the wideband uplink communication with \ac{snr} equal to $15$ dB, $\Nantennas = 8$, $B = 16$, and $Q = 4$.}\vspace{-0.5cm}
\label{fig:Rate_vs_N}
\end{figure}

\vspace{-0.3cm}
\subsection{Sum-Rate Performance}\vspace{-0.1cm}
We first examine in Fig$.$~\ref{fig:Rate_vs_SNR} the case where the sum rate varies with respect to the \ac{snr}, defined as $1/\sigma^2$, 
using $N = 20$ \ac{ris} unit elements. It can be seen that, for all evaluated schemes, the achievable sum-rate performances follow an increasing trend with increasing \ac{snr}. In addition, it is observed that, when using conventional \ac{ris} design methods which assume frequency-flat responses for wideband communications, one would expect to achieve superior rates (the red curve in Fig$.$~\ref{fig:Rate_vs_SNR}), while actually achieving much lower sum rates. These rates are notably outperformed by the proposed design which is aware of the underlying frequency selectivity. 
In Fig$.$~\ref{fig:Rate_vs_N}, we investigate the sum-rate performance versus the number $N$ of \ac{ris} elements, 
while fixing the \ac{snr} value to $15$ dB. As expected, the behavior of the achievable rate is increasing by employing more \ac{ris} unit elements, for all presented schemes. We again observe that configuring the \ac{ris}, assuming that its metamaterial elements realize frequency-flat phase shifters for wideband signals, yields performance which considerably deviates from that anticipated, and is notably degraded compared to our Lorentzian-response-aware RIS configuration design.

\vspace{-0.2cm}
\section{Conclusion}
\label{sec:concl}\vspace{-0.1cm}
In this paper, we studied the uplink of wideband \ac{ris}-empowered multi-user \ac{mimo} communication systems, where each \ac{ris} metametarial element operates as a resonant circuit, modeled according to a Lorentzian frequency response. We focused on the achievable sum-rate maximization problem, aiming to design the externally controlled oscillator strength, angular resonance frequency, and damping factor of each RIS element, instead of adjusting only its phase-shift angle. A BCD-based configuration algorithm was proposed to tackle the design of the frequency-selective \ac{ris}. Our simulation results demonstrated the sum-rate gains achievable using the proposed design compared to conventional configuration methods that assume frequency-flat metamaterials. In future work, we intend to compare the proposed frequency-selective modeling and optimization approaches with those in \cite{Li2021}.

\vspace{-0.2cm}
\bibliographystyle{IEEEtran}
\bibliography{IEEEabrv, references_edited}

\end{document}